\documentstyle[12pt,graphicx,caption2]{article}

\begin{document}

\title{Low-density limit as a test for the models of complicated
processes in the nuclei}
\author{V.I.Nazaruk\\
Institute for Nuclear Research of the Academy of Sciences of Russia,\\
60th October Anniversary Prospect 7a, 117312 Moscow, Russia.*}

\date{}
\maketitle
\bigskip

\begin{abstract}
We propose to use the branching ratio of channels of free-space hadron-nucleon
interaction as a test in the construction and verification of the models of
complicated processes in the nuclei. The particle self-energy and amplitudes
of subprocesses are studied as well.
\end{abstract}

\vspace{5mm}

\vspace{5mm}

\vspace{5cm}
*E-mail: nazaruk@inr.ru

\newpage

\setcounter{equation}{0}
Let us consider a free-space decay $a\rightarrow \pi ^0\bar{n}$, for example,
$\bar{\Lambda }\rightarrow \pi ^0\bar{n}$. For a decay in the medium the annihilation
\begin{equation}
a\rightarrow \pi ^0+\bar{n}\rightarrow \pi ^0+M
\end{equation}
and scattering
\begin{equation}
a\rightarrow \pi ^0+\bar{n}\rightarrow \pi ^0+\bar{n}
\end{equation}
channels take place. Here $M$ are the annihilation mesons, $\bar{n}\rightarrow M$
and $\bar{n}\rightarrow \bar{n}$ imply the annihilation and scattering of $\bar{n}$
in the medium, respectively. In [1,2] it was shown that the phenomenoligical model
based on optical potential does not describe the decay channel (1) as well as the
total decay probability of the $a$-particle. It describes the channel (2) only.
This is because the equation $\sum_{f\neq i}\mid T_{fi}\mid ^2\approx 2{\rm Im}T_{ii}$
cannot be used for the essentially non-unitary $S$-matrix. This is also true for the
more complicated decays and reactions [2].

In this connection we continue consideration of some aspects of unitary models of
multistep processes in the nuclear matter. Let $\Gamma _a$, $\Gamma _s$ and $\Gamma
_t$ be the widths of the decays (1), (2) and the total width of the decay $a\rightarrow
\pi ^0\bar{n}$ in the medium, respectively. (For a decay in nuclear matter $\Gamma
_s\approx 0$ since the $\bar{n}$ annihilates in a time $\tau _a\sim 1/\Gamma $, where 
$\Gamma $ is the annihilation width of $\bar{n}$ in the medium.) We calculate the 
$\Gamma _a$, $\Gamma _t$ and the branching ratio of channels. In the low-density 
approximation [3,4] the expressions for the branching ratio of channels and particle 
self-energy have a clear physical meaning (we emphasize this fact), which enables one 
to verify and correct the model. It is shown that the models of realistic processes 
in the nuclei should reproduce the branching ratio of channels of the corresponding 
free-space hadron-nucleon interactions. This can be considered as necessary condition 
for the correct model construction. The results obtained for the decay $a\rightarrow 
\pi ^0\bar{n}$ are generalized to the decay 
\begin{equation}
a\rightarrow b+c
\end{equation}
and reaction
\begin{equation}
a+N\rightarrow b+c
\end{equation}
followed by elastic and inelastic $c$-medium interactions. We focus on the
intermediate-state interaction of the $\bar{n}$ ($c$-particle) and so the $\pi
^0$-medium ($b$-medium) interaction is inessential for us. (In a recent paper [5],
the similar program has been realized for the $n\bar{n}$ transitions [6-8] in the 
medium followed by annihilation: $n\rightarrow \bar{n}\rightarrow M$. This is a 
simplest process involving the intermediate-state interaction since the $n\bar{n}$ 
transition vertex corresponds to 2-tail diagram. In this paper the consideration 
is generalized to the processes (3) and (4).)

We calculate the width of the decay (1). Our plan is as follows. 1) First of all
we construct and study the simplest process model sutable to the concrete calculations.
2) In the framework of this model we calculate the values listed above.

In order to construct the model correctly, we consider first the free-space $\bar{n}N$ 
annihilation (see Fig. 1a) and the process on a free nucleon
\begin{equation}
a+N\rightarrow \pi ^0+\bar{n}+N\rightarrow \pi ^0+M,
\end{equation}
shown in Fig. 1b (the free-space subprocess). The amplitude of free-space $\bar{n}N$
annihilation $M_a$ is defined as
\begin{equation}
<\!M\!\mid T\exp (-i\int dx{\cal H}_{\bar{n}N}(x))-1\mid\! \bar{n}N\!>=
N_a(2\pi )^4\delta ^4(p_f-p_i)M_a.
\end{equation}
Here ${\cal H}_{\bar{n}N}$ is the Hamiltonian of the $\bar{n}N$ interaction,
$N_a$ includes the normalization factors of the wave functions.
$M_a$ involves all the $\bar{n}N$ interactions followed by annihilation
including the {\em $\bar{n}N$ rescattering in the initial state}.

\begin{figure}[h]
  {\includegraphics[height=.25\textheight]{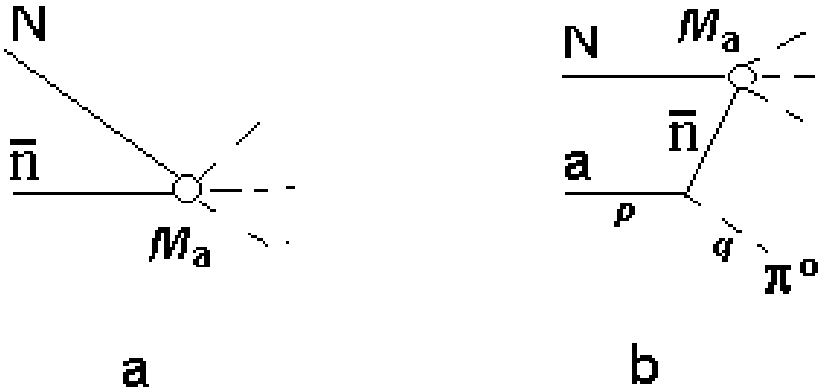}}
  \caption{{\bf a} Free-space $\bar{n}N$ annihilation. {\bf b} Free-space
reaction $a+N\rightarrow \pi ^0+\bar{n}+N\rightarrow \pi ^0+M$.}
\end{figure}

We write the formulas corresponding to Fig. 1b. The interaction Hamiltonian is
\begin{equation}
{\cal H}_I={\cal H}_d+{\cal H}_{\bar{n}N},
\end{equation}
\begin{equation}
{\cal H}_d=g\bar{\Psi }_{\bar{n}}\Phi ^* \Psi _a+H.c.,
\end{equation}
where ${\cal H}_d$ is the Hamiltonian of the decay $a\rightarrow \pi ^0\bar{n}$.
In the following the antineutron and $a$-particle are assumed non-relativistic
and spinless.

In the lowest order in ${\cal H}_d$ the amplitude of the process (5) is given by
\begin{equation}
M_{1b}=gGM_a,
\end{equation}
\begin{equation}
G=\frac{1}{(p_0-q_0-m)-({\bf p}-{\bf q})^2/2m+i0},
\end{equation}
where $p$ and $q$ are the 4-momenta of the $a$-particle and $\pi ^0$, respectively, 
$m$ is the antineutron mass, $M_a$ is given by (6). Since $M_a$ contains all the 
$\bar{n}N$ interactions followed by annihilation, the antineutron propagator $G$ is bare.

Consider now the decay (1) (see Fig. 2a). The background $a$-particle potential
is included in the wave function of $a$-particle $\Psi _a(x)$. The interaction
Hamiltonian has the form
\begin{equation}
{\cal H}_I={\cal H}_d+{\cal H},
\end{equation}
where ${\cal H}$ is the Hamiltonian of the $\bar{n}$-medium interaction. In the
lowest order in ${\cal H}_d$ the amplitude of the process, $M_{2a}$, is
uniquely determined by the Hamiltonian (11):

\begin{equation}
M_{2a}=gGM_a^m.
\end{equation}
The amplitude of the $\bar{n}$-medium annihilation $M_a^m$ is given by
\begin{equation}
<\!f\!\mid T\exp (-i\int dx{\cal H}(x))-1\mid\! 0\bar{n}_{p-q}\!>=
N(2\pi )^4\delta ^4(p_f-p_i)M_a^m
\end{equation}
(compare with (6)). Here $\mid\! 0\bar{n}_{p-q}\!>$ and $<\!f\!\mid $ are the
states of the medium containing the $\bar{n}$ with the 4-momentum $p-q$ and
annihilation products, respectively; $N$ includes the normalization factors of
the wave functions.

\begin{figure}[h]
  {\includegraphics[height=.25\textheight]{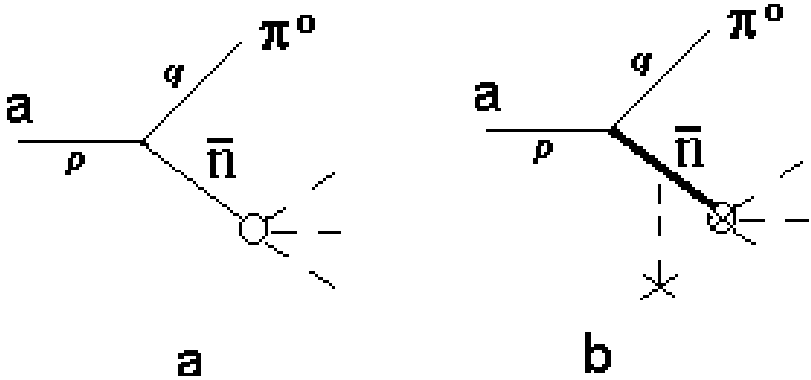}}
  \caption{{\bf a} Decay $a\rightarrow \pi ^0+\bar{n}\rightarrow \pi ^0+M$
in nuclear matter. The antineutron annihilation is shown by a circle. {\bf b}
The same as {\bf a}, but the antineutron propagator is dressed (see text).}
\end{figure}

The definition of the annihilation amplitude $M_a^m$ through Eqs. (13) is
natural. If the number of particles of medium is equal to unity, Eq. (13)
goes into (6). The antineutron annihilation width $\Gamma $ is expressed
through $M_a^m$ (see Eq. (20)). $M_a^m$ involves all the $\bar{n}$-medium
interactions followed by annihilation {\em including} the antineutron
rescattering in the initial state. Due to this, the propagator $G$ is bare.
As in (9), the antineutron self-energy $\Sigma =0$ since the interaction,
which can generate $\Sigma $, is involved in $M_a^m$.

If the Hamiltonian ${\cal H}$ is expressed through the $\bar{n}N$- and
$NN$-interactions, the amplitude $M_a^m$ contains in medium $\bar{n}N$-amplitudes
and dressed propagators. In this case the following condition should be fulfilled:
if $\rho \rightarrow 0$, the propagator is not dressed. However, our purpose is to
study the general features of the simplest model {\em suitable to the concrete
calculations} and not the medium effects. On this reason we consider the model (12)
which contains the block $M_a^m$ corresponding to the observable values. We would
like to emphasize this paragraph.

Our results (Eqs. (29)-(35)) may not depend on the model of the block $M_a^m$.
However, for the concrete calculations given in this paper, the fact that the
propagator is bare is principal and so we study this point in detail. Using the 
same Hamiltonian (11), we try to construct the model with the dressed propagator
(see Fig. 2b). Let $U_{\bar{n}}$ be the optical potential of $\bar{n}$. Denote
\begin{equation}
V={\rm Re}U_{\bar{n}}.
\end{equation}
We recall the Hamiltonian ${\cal H}$ involves all the $\bar{n}$-medium interactions.
In the Hamiltonian ${\cal H}$ we separate out the real potential $V$:
\begin{equation}
{\cal H}=V\bar{\Psi }_{\bar{n}}\Psi _{\bar{n}}+{\cal H}'
\end{equation}
and include it in the antineutron Green function
\begin{equation}
G_d=G+GVG+...=\frac{1}{G^{-1}-V}.
\end{equation}
The amplitude $M_{2b}$ of the process shown in Fig. 2b is
\begin{equation}
M_{2b}=gG_dM'_a.
\end{equation}
Since the amplitudes $M_{2a}$ and $M_{2b}$ correspond to one and the
same Hamiltonian (11), $M_{2a}=M_{2b}$ and
\begin{equation}
G_dM'_a =GM_a^m.
\end{equation}
The propagator $G_d$ is dressed: $\Sigma =V\neq 0$. According to (18), the
expressions for the propagator and vertex function are uniquely connected
(if $H_I$ is fixed). The "amplitude" $M'_a(V,H')$ should describe the
annihilation. However, below is shown $M'_a$ and model (17) are unphysical.
Comparing the left- and right-hand sides of (18), we see the following.

(1) If the number of particles of medium $n$ is equal to unity, model (17)
does not describe the free-space process shown in Fig. 1b because Eq. (9)
contains the bare propagator.

(2) The observable values ($\Gamma $, for example) are expressed through $M_a^m$
and not $M'_a$. Compared to $M_a^m$, $M'_a$ is truncated because the portion of
the Hamiltonian $H$ is included in the $G_d$. $M'_a$ has not a physical meaning.

(3) Amplitude (17) cannot be naturally obtained from the formal expansion of the
$T$-operator $T\exp (-i\int dx(V\bar{\Psi }_{\bar{n}}\Psi _{\bar{n}}+{\cal H}'))$.

(The formal expression for the dressed propagator should contain the annihilation
loops as well. In this case the statements given in pp. (1) and (2) are only
enhanced. A particle self-energy should be considered in the {\em context of
the concrete problem}. The dressed propagator arises naturally if $V$ and
${\cal H}'$ are the principally different interactions and vertex function
does not depend on $V$. In our problem one and the same field generates
$\Sigma $ and $M_a^m$.)

(4) Equations (16) and (17) mean that the annihilation is turned on upon forming
of the self-energy part $\Sigma =V$ (after multiple rescattering of $\bar{n}$).
This is counter-intuitive since at the low energies [9-11]
\begin{equation}
\frac{\sigma _a}{\sigma _t}>0.7
\end{equation}
($\sigma _t=\sigma _a+\sigma _s$, $\sigma _a$ and $\sigma _s$ are the cross
sections of free-space $\bar{n}N$ annihilation and scattering, respectively)
and inverse picture takes place: in the first stage of $\bar{n}$-medium
interaction the annihilation occurs.

The realistic {\em competition} between scattering and annihilation should
be taken into account. Both scattering and annihilation vertices should
occur on equal terms in $M_a^m$ or $G_d$. According to pp. (1)-(3) the
latest possibility should be excluded. Model (12) is free from
drawbacks given in pp. (1)-(3). It reproduces the ratio (19) as well.

\begin{figure}[h]
  {\includegraphics[height=.25\textheight]{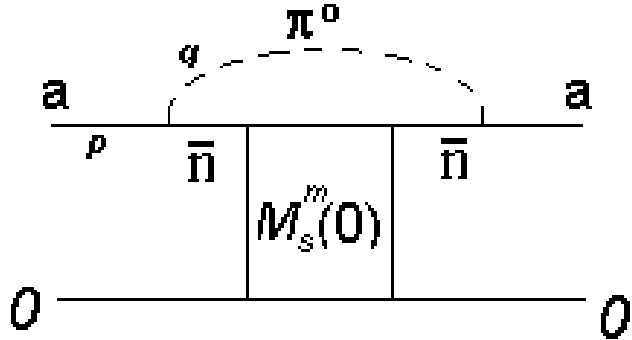}}
  \caption{On-diagonal matrix elements corresponding to the process
$(a-\mbox{medium})\rightarrow \pi ^0+(\bar{n}-\mbox{medium})\rightarrow
(a-\mbox{medium})$.}
\end{figure}

Indeed, we calculate the width of the decay (1) shown in Fig. 2a. Put ${\bf p}=0$
for simplicity. The pion wave function is $\Phi (x)=(2q_0\Omega )^{-1/2}\exp (-iqx)$.
Using the amplitude (12), the decay width $\Gamma _a$ is found to be
\begin{equation}
\Gamma _a=\int d{\bf q}\frac{1}{2q_0(2\pi )^3}g^2G^2\Gamma ,
\end{equation}
where $\Gamma $ is the annihilation rate of the $\bar{n}$ with the 4-momentum $p-q$.
The corresponding amplitude $M_a^m$ is given by (13).

Up to this point the low-density limit has been not used. The number of exchanges
with the medium or particle, which appears in (16), is very important but
unobservable value. In particular, it is responsible for the particle self-energy
and process suppression by the potential $V$. The low-density approximation enables
one to verify directly the condition (19). Indeed, in the low-density
approximation $\Gamma =v\rho \sigma _a$ and
\begin{equation}
\Gamma _a=\int d{\bf q}F_1\sigma _a,
\end{equation}
\begin{equation}
F_1=\frac{g^2G^2v\rho }{2q_0(2\pi )^3},
\end{equation}
$q_0^2={\bf q}^2+m_{\pi }^2$, where $m_{\pi }$ is the pion mass.

Let us calculate the total width $\Gamma _t$ of the decay $a\rightarrow \pi ^0\bar{n}$
in the medium. In the lowest order in ${\cal H}_d$ the on-diagonal matrix
element shown in Fig. 3 is given by
\begin{equation}
M_{3}=\int \frac{dq}{(2\pi )^4}g\frac{i}{q^2-m_{\pi }^2+i0}GM_s^m(0)Gg.
\end{equation}
Here $M_s^m(0)$ is the forward scattering amplitude of $\bar{n}$ in the medium.
We integrate over $q_0$ and use the optical theorem in the left- and right-hand
sides of (23):
\begin{equation}
\frac{1}{T_0}2{\rm Im}M_{3}=\Gamma _t,
\end{equation}
\begin{equation}
\frac{1}{T_0}2{\rm Im}M^m_s(0)=v\rho \sigma _t
\end{equation}
($T_0$ is the normalization time, $T_0\rightarrow \infty  $), resulting in
\begin{equation}
\Gamma _t=\int d{\bf q}F_1\sigma _t.
\end{equation}
Denoting $\mid\!{\bf q}\!\mid =q$, one gets finally
\begin{equation}
r=\frac{\Gamma _a}{\Gamma _t}=\frac{\int dqF(q)\sigma _a(q)}{\int dqF(q)\sigma
_t(q)},
\end{equation}
\begin{equation}
F=\frac{q^3G^2}{q_0},
\end{equation}
$q_0^2={\bf q}^2+m_{\pi }^2$. In the differential form
\begin{equation}
\frac{d\Gamma _a/dq}{d\Gamma _t/dq}=\frac{\sigma _a}{\sigma _t}.
\end{equation}

In a similar manner, one obtains
\begin{equation}
\frac{d\Gamma _s/dq}{d\Gamma _t/dq}=\frac{\sigma _s}{\sigma _t}.
\end{equation}

We consider the free-space decay (3). Let for the decay in the medium the
inelastic
\begin{equation}
a\rightarrow b+c\rightarrow b+x
\end{equation}
and elastic
\begin{equation}
a\rightarrow b+c\rightarrow b+c
\end{equation}
interactions of the $c$-particle take place. Here $c\rightarrow x$ and $c\rightarrow
c$ imply the reaction induced by $c$-particle and scattering of $c$-particle in the
medium, respectively. Let $\Gamma _r$ and $\Gamma _s$ be the widths of decays (31)
and (32), respectively; $\Gamma _t=\Gamma _r+\Gamma _s$. With the replacement
$\Gamma _a\rightarrow \Gamma _r$ and $\sigma _a\rightarrow \sigma _r$, the relations 
(27)-(30) describe the decay channels (31) and (32).

For the process (4) in the medium instead of (31) and (32) we consider the reactions
on the nucleon of the medium:
\begin{equation}
a+N\rightarrow b+c\rightarrow b+x
\end{equation}
and
\begin{equation}
a+N\rightarrow b+c\rightarrow b+c.
\end{equation}
In this case the similar relations take place as well. The corresponding changes in
$F$ are minimum and non-principal. For example, for the channels (33) and (34) we get
\begin{equation}
\frac{d\sigma _r^m/dq}{d\sigma _s^m/dq}=\frac{\sigma _r}{\sigma _s},
\end{equation}
where $\sigma _r^m$ and $\sigma _s^m$ are the cross sections of the reactions
(33) and (34), respectively.

Relations (27)-(30) are directly connected with the experimental branching ratios 
$\sigma _a/\sigma _t$ and $\sigma _s/\sigma _t$ (see (19)). The similar statements 
are also true for the decays like (31), (32) and reactions (33), (34) in the
medium. The nuclear medium changes the amplitudes [12] and branching ratio of channels
[13] in hadron-nucleon interactions. Besides, the strong antineutron absorption takes
place. Due to this, at the nuclear densities $\Gamma _a/\Gamma _t\approx 1$. However,
if $\rho \rightarrow 0$, the relations (27)-(30) should be reproduced.

The definition of annihilation amplitude through Eqs. (6) and (13) is natural since 
it corresponds to the observable values. In that event the intermediate particle 
propagator is bare and relation (19) is reproduced. (This does not contradict to 
well-known results [14-18] because $\Sigma $ should be considered in the context of 
the concrete problem.) Certainly, we do not argue that above-considered scheme is the 
only possible model. This is clear even from Eq. (18). Besides, the Hamiltonian 
${\cal H}$ can be concretized. We argue only the following: once the amplitudes are 
defined by (6) and (13) (which is natural), the propagator is bare; the free-space 
processes shown in Fig. 1 and the experimental branching ratio of channels like (19) 
are reproduced. In fact, the relations like (29), (30) and (35) should be fulfilled 
for any process model and can be considered as necessary condition for the correct 
model construction.

\end{document}